\documentclass[apjl,a4paper,12pt,useAMS]{emulateapj}
\usepackage{amsmath}
\usepackage{cases}
\usepackage{amssymb}
\usepackage{graphicx}

\begin{document}
\title{What powered the optical transient AT2017gfo associated with GW170817?}
\author{Shao-Ze~Li$^{1,2}$, Liang-Duan~Liu$^{2,3,5}$, Yun-Wei~Yu$^{1,4}$, and Bing~Zhang$^{2}$}

\altaffiltext{1}{Institute of Astrophysics, Central China Normal
University, Wuhan 430079, China, {yuyw@mail.ccnu.edu.cn}}
\altaffiltext{2}{Department of Physics and Astronomy, University of
Nevada Las Vegas, Las Vegas, NV 89154, USA,
{zhang@physics.unlv.edu}} \altaffiltext{3}{School of Astronomy and
Space Science, Nanjing University, Nanjing 210093, China}
\altaffiltext{4}{Key Laboratory of Quark and Lepton Physics (Central
China Normal University), Ministry of Education, Wuhan 430079,
China} \altaffiltext{5}{Key Laboratory of Modern Astronomy and
Astrophysics (Nanjing University), Ministry of Education, Nanjing
210093, China}

\begin{abstract}
The groundbreaking discovery of the optical transient AT2017gfo
associated with GW170817 opens a unique opportunity to study the
physics of double neutron star (NS) mergers. We argue that the
standard interpretation of AT2017gfo as being powered by radioactive
decays of r-process elements faces the challenge of simultaneously
accounting for the peak luminosity and peak time of the event, as it
is not easy to achieve the required high mass, and especially the
low opacity of the ejecta required to fit the data. A plausible
solution would be to invoke an additional energy source, which is
probably provided by the merger product. We consider energy
injection from two types of the merger products: (1) a post-merger
black hole powered by fallback accretion; and (2) a long-lived NS
remnant. The former case can only account for the early emission of
AT2017gfo, with the late emission still powered by radioactive
decay. In the latter case, both early- and late-emission components
can be well interpreted as due to energy injection from a
spinning-down NS, with the required mass and opacity of the ejecta
components well consistent with known numerical simulation results.
We suggest that there is a strong indication that the merger product
of GW170817 is a long-lived (supramassive or even permanently
stable), low magnetic field NS. The result provides a stringent
constraint on the equations of state of NSs.
\end{abstract}

\section{introduction}
The discovery of the first gravitational wave (GW) event, i.e.
GW150914 from a merger of double black holes (BHs), marked the
beginning of the era of GW astronomy (Abbott et al. 2016). On 2017
August 17,  the Laser Interferometer Gravitational-Wave Observatory
(LIGO)/Virgo detector network further detected a historical event
GW170817, the first GW event from the merger of a neutron
star-neutron star (NS-NS) binary (Abbott et al. 2017a), which was
followed by a short-duration gamma-ray burst (GRB) dubbed GRB
170817A, captured by the Fermi satellite 1.7 s after the GW merger
event (Goldstein et al. 2017; Savchenko et al. 2017; Zhang et al.
2018). The GW170817/GRB 170817A association robustly confirmed the
long-standing hypothesis that short GRBs originate from compact star
mergers involving at least one NS. The apparently low radiation
luminosity and energy of GRB 170817A are consistent with having this
GRB being observed at a large viewing angle from the jet axis
(Abbott et al. 2017b), which causes the missing afterglow emission
during the first $\sim10$ days in follow-up observations (Troja et
al. 2017). Nevertheless, during this period, a significant
ultraviolet-optical-infrared (UVOIR) transient was detected, first
announced by Coulter et al. (2017) and subsequently observed by many
groups (e.g., Arcavi et al. 2017; Lipunov et al. 2017; Tanvir et al.
2017; Valenti et al. 2017). This transient, named as
AT2017gfo/SSS17a/DLT17ck (hereafter AT2017gfo), was thought to be
associated with an NS-NS merger (Li \& Paczy\'{n}ski 1998), which
has been called a ``kilonova'' (Metzger et al. 2010) or a
``mergernova'' (Yu et al. 2013; Li \& Yu 2016). In the rest of the
paper, we use the term ``mergernova'' for the following two reasons:
(1) a ``kilonova'' is defined as being powered by radioactive decay.
As shown below, we invoke energy injection from a central engine to
account for the observations. The term ``mergernova'' broadly
defines the merger-associated UVOIR transients, regardless of the
energy power. (2) The reason for adopting the kilonova terminology
was that the peak luminosity is about 1000 times of that of a
typical nova, which is $10^{41} \ {\rm erg \ s^{-1}}$. The earliest
observational data point of AT2017gfo already has a luminosity above
$10^{42} \ {\rm erg \ s^{-1}}$, at least one order of magnitude
brighter than the typical luminosity of kilonovae. As shown by Yu et
al. (2013) and Metzger \& Piro (2014), the existence of a long-lived
NS as the post-merger product can increase the peak luminosity
significantly. A BH central engine but with additional accretion
activities may also act as a source of energy injection to the
mergernova (Ma et al. 2018; Song et al. 2018).

The property of a radioactivity-powered mergernova primarily depends
on the mass and the opacity of the ejecta. In particular, the
existence of lanthanides, even with a small mass fraction (e.g.
$\sim10^{-4}$), would greatly increase the Planck mean opacity by as
much as $\sim(10-100)\,\rm cm^{2}g^{-1}$ (Kasen et al. 2013), so
that the peak time of the event will be shifted to about one week
after the merger, with a redder spectrum at the peak (Barnes \&
Kasen 2013; Tanaka \& Hotokezaka 2013). In any case, a polar
outflow, most likely launched by a disk wind and irradiated by
neutrino emission, may still give rise to an early ``blue''
component because lanthanide synthesis is probably inefficient there
(Metzger \& Fern\'andez 2014). The observed AT2017gfo emission can
be indeed understood with such a ``blue+red'' radioactivity-powered
mergernova model (e.g., Cowperthwaite et al. 2017; Nicholl et al.
2017; Smartt et al. 2017; Tanaka et al. 2017; Villar et al.
2017).\footnote{In some papers, the dynamical ejecta is interpreted
as ``blue'', whereas the disk wind outflow as ``red'' (e.g. Kasen et
al. 2017).} Specifically, the interpretation of the peak luminosity
$\sim10^{42} \, \rm erg\,s^{-1}$ and the peak time $\sim$ 1 day in
this model requires a relatively low opacity ($\kappa\sim0.3 \, \rm
cm^{2}g^{-1}$), a relatively large ejecta mass
($M\sim0.04\,M_{\odot}$), and a relatively high characteristic
velocity ($v\sim0.3c$), for the blue component. These requirements
push the boundary of numerical simulations regarding the ejected
mass (Dessart et al. 2009; Fern\'{a}ndez \& Metzger 2013; Perego et
al. 2014; Just et al. 2015; Richers et al. 2015; Shibata et al.
2017) and the expected opacity, which is believed to be, at least,
not much lower than $\sim 1\rm cm^2g^{-1}$ (Kasen et al. 2013;
Tanaka et al. 2018).

Metzger et al. (2018) argued that a short-lived hypermassive NS with
a surface magnetic field of $B\sim10^{14}\,\rm G$ could help to
increase the mass of a disk wind. Radice et al. (2018) also
suggested that the viscous ejecta can be as much as $0.1 M_{\odot}$.
These can partially decrease the difficulty of the radioactive
mergernova model, but the required low opacity may not be readily
accounted for. Alternatively, if the remnant NS is long-lived, then
the mergernova emission itself would be significantly affected by
the NS due to the additional energy injection from the NS and the
effect of ionization (Yu et al. 2013; Metzger \& Piro 2014).
Recently, Yu et al. (2018) showed that the observed emission from
AT2017gfo can be accounted for by a hybrid model, with the early
emission powered by radioactivity and later emission powered by
energy injection from a long-lived, low-field pulsar\footnote{The
relatively low luminosity of the prompt emission and broadband
afterglow of GRB 170817A has also significantly constrained the
properties of a putative underlying NS, with a dipolar magnetic
field that should be significantly below the range of a typical
magnetar (Ai et al. 2018; Geng et al. 2018).}. In their modeling,
all of the emission comes from the same ejecta component, with a
single uniform opacity around $1\rm cm^2g^{-1}$ and a total mass of
$\sim0.03M_{\odot}$. The latter is less than a half of the total
mass invoked to interpret the event using the radioactive heating
alone (e.g., $0.065M_{\odot}$; Villar et al. 2017). It remains
unclear whether ionization by the pulsar wind can penetrate deep
enough to reduce the opacity of the entire ejecta to be around
$\sim1\rm cm^{2}g^{-1}$.

This {\em Letter} includes two parts. The first part (Section 2)
presents an argument against the traditional radioactivity-powered
mergernova model; specifically, the difficulty in simultaneously
accounting for both the high luminosity (high mass) and early peak
time (low opacity). Encouraged by Yu et al. (2018), the second part
(Section 3) presents our modeling of AT2017gfo within the framework
of engine-powered mergernova. We show that without an ad hoc
mechanism to drastically reduce the opacity of the merger ejecta and
with a reasonable amount of ejecta mass (a few $10^{-3} M_\odot$),
the observational data can be accounted for, given that the merger
left behind a long-lived, low-$B$ NS.

\section{The radioactivity power}\label{mergernova}
\begin{figure}
\centering\resizebox{\hsize}{!}{\includegraphics{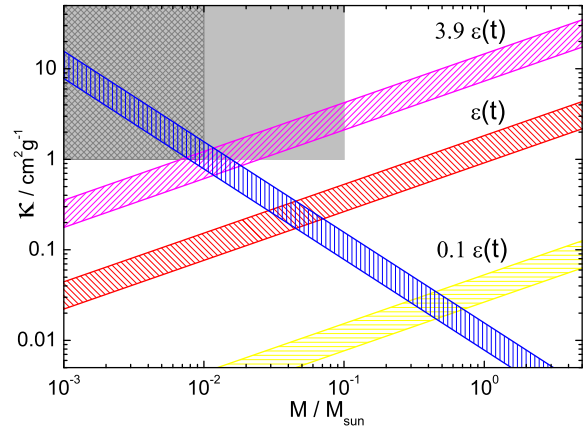}}
\caption{Constraints on the opacity $\kappa$ and ejecta mass $M$ for
the blue component of AT2017gfo. The constraints from the peak time
and peak luminosity are shown in the blue and red stripes,
respectively, with the velocity constrained to $(0.3 \pm 0.1) c$.
The preferred parameter regime based on numerical simulations  and
theoretical calculations is marked as gray area at the upper-left
corner. The uniform+striped gray area and the striped gray area are
for the dynamical ejecta and disk wind, respectively. Considering
inefficient heating due to a large $Y_{\rm e}$ (favorable for a low
$\kappa$) would reduce the heating rate by one order of magnitude
(lower yellow stripe). Fixing $\kappa \sim 1 \,{\rm cm^2 g^{-1}}$,
the required heating rate is larger than what radioactive heating
can provide (upper magenta stripe), suggesting the existence of a
central engine heating source. } \label{parameter constraints}
\end{figure}

In the traditional radioactivity-powered mergernova (kilonova)
model, the parameters of the blue-component ejecta (the polar disk
wind) can be estimated from the peak of the observational bolometric light curve of AT2017gfo by the following analytical
method. According to Arnett (1982), the peak bolometric luminosity
is equal to the heating rate at the peak time, i.e.,
\begin{eqnarray}
L_{\rm p}&=&\eta_{\rm th}M\varepsilon(t_{\rm p})\label{Lp},
\end{eqnarray}
where $\eta_{\rm th}$ is the thermalization efficiency, $M$ is the
mass and $\varepsilon$ is the heating rate per unit mass. The
heating rate firstly remains constant for a duration of about one
second and then decays following a power law that can be roughly
estimated as (Metzger et al. 2010; Korobkin et al. 2012)
\begin{eqnarray}
\varepsilon(t)\approx2\times10^{10}{\rm erg\,g^{-1}s^{-1}}
\left({t\over1\,\rm day}\right)^{-\alpha}\label{heating},
\end{eqnarray}
where $\alpha=1.3$. One can further use the characteristic diffusion
timescale of the ejecta to estimate the peak emission time, which
reads (Metzger 2017)
\begin{eqnarray}
t_{\rm p}&=&\left(3M\kappa\over4\pi \beta v c\right)^{1/2}\nonumber\\
&\approx&1.6\,{\rm d}\left({M\over
0.01M_{\odot}}\right)^{1/2}\left({v\over
0.1\,c}\right)^{-1/2}\left({\kappa\over 1\,{\rm
cm^2\,g^{-1}}}\right)^{1/2},\label{tpeak}
\end{eqnarray}
where $\beta=3$ is a dimensionless parameter characterized by the
density profile of the ejecta. As a result, the peak
luminosity can be determined to
\begin{eqnarray}
L_{\rm p}&=&1.2\times 10^{41}{\rm erg\,s^{-1}}\,\nonumber\\
&&\left({M\over 0.01M_{\odot}}\right)^{1-\alpha/2}\left({v\over
0.1\,c}\right)^{\alpha/2}\left({\kappa\over 1\,{\rm
cm^2\,g^{-1}}}\right)^{-\alpha/2}\label{Lpeak},
\end{eqnarray}
where the thermalization efficiency is adopted as $\eta_{\rm
th}\sim0.5$ following Barnes et al. (2016). By taking the
observational peak values of $L_{\rm p}\sim10^{42}\rm erg~s^{-1}$
and $t_{\rm p}\sim1$ day, we can constrain $M$ and $\kappa$ from
Equations (\ref{tpeak}) and (\ref{Lpeak}) given a range of allowed
ejecta velocity. The high $L_{\rm p}$ demands large $M$, large $v$
and small $\kappa$. In order to get a small $t_{\rm p}$, again a
large $v$ and small $\kappa$ is preferred. We adopt a relatively
large velocity $v\sim 0.3c$, but allow a range of $\pm 0.1c$ in our
discussion. The required parameters are centered around $M\sim
0.04\rm M_{\odot}$ and $\kappa\sim0.3\rm cm^2g^{-1}$, as shown in
Figure \ref{parameter constraints}. Waxman et al.(2017) also gave a
constraint about the required opacity under the
radioactivity-powered model, which should be at least $\lesssim
0.3\rm cm^2g^{-1}$. For comparison, we also present the allowed
values of the parameters $M_{\rm ej}$ and $\kappa$ in Figure
\ref{parameter constraints}, as shown by the gray area, where the
upper limit on the ejecta mass is taken as $M<0.1M_{\odot}$ for the
dynamical ejecta (uniform+striped gray area) and $M<0.01M_{\odot}$
for the disk wind (striped gray area; Bauswein et al. 2013;
Hotokezaka et al. 2013; Rosswog 2013; Just et al. 2015; Richers et
al. 2015; Shibata et al. 2017; Siege and Metzger et al. 2018). The
lower limit on opacity is taken as $\kappa>1\,\rm cm^{2}g^{-1}$ by
considering the high velocity of the merger ejecta (Kasen et al.
2013; Tanaka et al. 2018). It is clearly shown that the required
parameters by the radioactivity-powered mergernova model are far
away from the allowed parameter regions. This makes this model a
disfavored one.

More specifically, although $M\sim 0.04\rm M_{\odot}$ is already too
high for a disk wind, it could still be acceptable if the merger
product is a highly magnetized NS (Metzger et al. 2018). The more
serious issue comes from the small opacity $\kappa\sim0.3\rm
cm^2g^{-1}$ demanded by the data. Detailed studies showed that the
opacity depends on the electron fraction $Y_{\rm e}$, which defines
how ``neutron rich'' the ejecta is. According to these studies, the
$r$-process reactions are only efficient for an electron fraction of
$Y_{\rm e}\lesssim 0.25$, in which case a remarkable number of heavy
elements of a mass number $A>130$ can be synthesized (Kasen et al.
2015; Rosswog et al. 2017). However, in the polar direction, the
electron faction of a disk wind is probably higher than 0.25 due to
the irradiation by the neutrino emission from the disk and,
sometimes, from a remnant NS. Specifically, the electron fraction
could be within the range of $Y_{\rm e}\sim 0.2-0.4$ if the remnant
is a promptly formed BH (Fern\'{a}ndez \& Metzger 2013;
Fern\'{a}ndez et al. 2015) or $Y_{\rm e}\sim0.3-0.5$ if the remnant
is a short-lived hypermassive NS (Metzger \& Fern\'{a}ndez 2014;
Metzger et al. 2018). Therefore, as the first impression, the low
opacity of $\kappa \sim0.3\rm cm^2g^{-1}$ for the peak emission of
AT2017gfo seems reasonable, if the remnant NS can live for a short
time. However, we would like to point out that the opacity can
actually be increased significantly due to the Doppler broadening of
bound-bound transitions (Karp et al. 1977), if the material has a
very high-velocity gradient which is indeed the situation in a
merger ejecta. Specifically, the Doppler effect due to the velocity
gradient can force the photons, with energies that do not strictly
match the energy-level differences, to be absorbed, which is
forbidden in the laboratory. Therefore, this so-called expansion
opacity is dependent on the distributions of density, temperature,
and velocity gradient of the ejecta. For a homogenous explosion, the
velocity gradient depends on the maximum velocity of the ejecta. In
a SN Ia ejecta with a velocity that is about several thousands of
$\rm km\, s^{-1}$, its typical opacity is on the order of $0.1\,\rm
cm^{2}g^{-1}$ for Fe-peak elements (Pinto \& Eastman 2000). In
contrast, an NS-NS merger typically launches with much greater
velocities, reaching a significant fraction of the speed of light.
The opacity of such merger ejecta can easily reach $1\,\rm
cm^{2}g^{-1}$, even if only the contributions from open d-shell
elements (i.e., Fe, Co, Ni, Ru, et al.) are considered and the
effects of lanthanides are ignored (Kasen et al. 2013; Tanaka et al.
2018). In other words, the value of $\sim 1\,\rm cm^{2}g^{-1}$ gives
a conservative lower limit of the opacity of the merger ejecta,
which has been widely adopted for lanthanide-free ejecta in the
studies before the detection of AT2017gfo (see reviews by
Fern\'{a}ndez \& Metzger 2016; Metzger 2017). As a result, we
believe that the low opacity of $\kappa \sim0.3\,\rm cm^{2}g^{-1}$
required by the radioactivity-powered mergernova model poses a great
challenge to the radioactivity-powered mergernova (or kilonova)
model.

Making things worse, for $Y_{\rm e}> 0.25$ that corresponds a low
opacity, not only is the lanthanides synthesis is blocked, but the
synthesis of other heavy elements could also be suppressed
significantly. In other words, the radioactive heating rate
decreases with an increasing electron fraction (Grossman et al.
2014; Wanajo et al. 2014; Lippuner \& Roberts 2015). Specifically,
the heating rate as presented in Equation (\ref{heating}) may be
relevant for $Y_{\rm e}\sim(0.1-0.3)$ but would start to decrease as
the electron fraction becomes larger than $\sim0.3$. For a
relatively high $Y_{\rm e}\sim(0.4-0.5)$ that is favored by a low
opacity, the heating rate could be reduced by one order of
magnitude. This further raises the required ejecta mass (e.g. $\sim
0.4 M_{\odot}$) to the unacceptable range.

In summary, the radioactivity-powered mergernova model faces a great
challenge, if it is not completely ruled out. As a possible solution
to these difficulties, an extra heating source is needed. Such a
heating power cannot be provided by the radioactive decay of heavy
elements, but can be provided by an underlying engine.

\section{Engine-powered mergernova model}\label{injection}

The chirp mass of the progenitor binary of GW170817 was derived to
$M_{\rm c}=1.188^{+0.004}_{-0.002}\rm M_{\odot}$ from LIGO
observations. This constrains the individual masses of the component
NSs to be in the range of $1.17-1.6\rm M_{\odot}$ by assuming low
spins for the NSs  and the total gravitational mass of the binary to
be about $2.74\rm M_{\odot}$ (Abbott et al. 2017a). After the GW
chirp and mass ejection, the gravitational mass of the remnant
object could be around $M_{\rm RNS}\sim2.6\rm M_{\odot}$ (Ai et al.
2018; Banik \& Bandyopadhyay 2017). The nature of this remnant is
subject to debate because of the uncertainties of an NS equation of
state. Let us denote the maximum mass of a non-rotating NS by
$M_{\rm TOV}$ and the maximum NS mass of a maximally rotating NS as
$M_{\rm max}$. The remnant would collapse into a BH promptly, or
after a brief hypermassive NS phase, if $M_{\rm max}<M_{\rm RNS}$.
For $M_{\rm TOV}<M_{\rm RNS}< M_{\rm max}$, the NS is supramassive
and can survive for an extended period of time until centrifugal
support can no longer hold against gravity. If $M_{\rm TOV} > M_{\rm
RNS}$, then the NS can live permanently. Depending on the outcome of
the remnant, the central engine can give rise to different
properties of energy injection to power the mergernova.

In the following, we discuss two types of central engines within the framework of the
engine-driven mergernova model (Yu et al. 2013, see Appendix).

\begin{figure}
\centering\resizebox{\hsize}{!}{\includegraphics{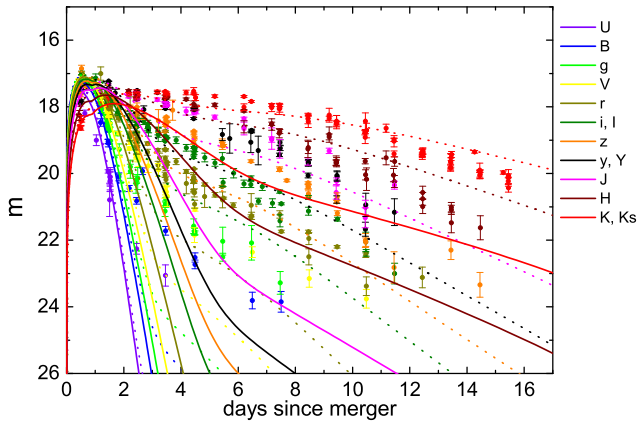}}
\caption{Fitting to the multiband light curves of AT2017gfo with the
BH fallback accretion engine. The fitting parameters are shown in
Table \ref{table}. The solid curves are for central engine-powered
emission only. The dotted curves include the contribution from
radioactive decay. The data are taken from Villar et al. (2017). The
distance is adopted as $D=40\rm Mpc$. }\label{fitfb}
\end{figure}

\begin{table*} \centering \caption{Fitting parameters}
\renewcommand{\arraystretch}{2.0}

\begin{tabular}{llllllllllllll}
 \hline
  \hline
    & $M_{\rm fb,i}/M_{\odot}\rm s^{-1}$ & $L_{\rm md,i}/\rm erg\,s^{-1}$ & $t_{\rm sd,gw}/\rm\,s$
  & $B/G$ & $\epsilon$ &
    Ejecta
    & $M_{\rm ej}/M_{\odot}$ & $\kappa/\rm cm^{2}g^{-1}$ & $v_{\rm ej,i}/c$ & $\Omega$ & $\delta$ & $\zeta$ & $A$ \\
    \hline
 Fallback    & $3\times10^{-3}$& - &-&-& -& Polar & $3\times10^{-3}$ & 1 & 0.25 & $2\pi$ & -1& 10& 6\\
                   &  &  &&&    & Equatorial & $5\times 10^{-2}$ & 5 & 0.15& $2\pi$ & -1& 10& 6\\
    \hline
  NS    & - & $3.4\times 10^{44}$ & $500$& $3.4\times10^{12}$ & 0.0035  &  Polar & $1\times10^{-3}$ & 1 & 0.35& $2\pi$ & -1& 10& 6\\
                   &  &  &  & & & Equatorial & $5\times 10^{-3}$ & 5 & 0.2& $2\pi$ & -1& 10& 6\\
\hline \hline
\end{tabular}\label{table}
\end{table*}

\subsection{BH with fallback accretion}
We first consider the case that the merger product of GW170817 is a
BH (including a prompt BH or a BH formed after a brief hypermassive
NS phase). In this case,  energy injection from the remnant BH could
only be due to fallback accretion. Although most of the ejecta would
be unbound, there is still a fraction of mass that is
gravitationally bound and would fallback onto the BH during a range
of time scales (Rosswog 2007). Initially, the fallback accretion
rate keeps constant over a small period of time $\sim0.1\,\rm s$,
and then decays following a power law $\propto\,t^{-5/3}$. The
heating rate to a mergernova due to this accretion may be calculated
by assuming that it is proportional to the accretion rate. We then
have (Metzger 2017)
\begin{eqnarray}
L_{\rm fb}&=&\eta_{\rm fb}\dot{M}_{\rm fb}c^{2}\\
&\approx&2\times10^{51}{\rm erg\,s^{-1}}\left({\eta_{\rm
fb}\over0.1}\right)\left({\dot{M}_{\rm
fb,i}\over10^{-3}M_{\odot}\,\rm s^{-1}}\right)\left(t\over
0.1\rm\,s\right)^{-5/3},
\end{eqnarray}
where $\dot{M}_{\rm fb}$ is the accretion rate with the subscript
`i' standing for ``initial'', and $\eta_{\rm fb}$ represents the
fraction of the accretion energy that can be ejected outwards (i.e.,
accretion feedback efficiency).

Tentative fitting with such an accretion-induced heating rate to the
multiband light curves of AT2017gfo is presented in Figure
\ref{fitfb}. Without a radioactive power included (solid curves),
the model can only account for the early blue component. In this
fitting, the opacity is fixed to $\kappa=1\,\rm cm^{2}g^{-1}$ for
the polar blue ejecta and $\kappa=5\,\rm cm^{2}g^{-1}$ for the
equatorial ejecta, corresponding to the lanthanide-free and
lanthanide-rich ejecta, respectively. The values of the other
parameters are taken freely and their values are presented in Table
1. For the adopted feedback efficiency of $\eta_{\rm fb}=0.1$
(Metzger 2017), the initial accretion rate is required to be $\dot
M_{\rm fb,i}=3\times 10^{-3}\,\dot{M}_{\odot}\rm s^{-1}$, which
corresponds to a total mass of $7.5\times 10^{-4}\,M_{\odot}$ of the
fallback material. In principle, the feedback efficiency $\eta_{\rm
fb}$ can be (much) smaller than 0.1, which would lead to a
requirement of a much higher and even unacceptable fallback
accretion rate. In any case, because the accretion heating decays
very quickly ($\propto t^{-5/3}$), the late-time emission of
AT2017gfo always needs to be powered by radioactive decay so that
the required ejecta mass is still substantial (e.g. $\sim0.05\rm
M_{\odot}$; Villar et al. 2017). According to numerical simulations,
such a high ejecta mass is only available for BH-NS mergers (Foucart
et al. 2013), but not for an NS-NS merger like GW170817.
 We therefore conclude that energy injection due to BH fallback accretion cannot satisfactorily interpret the data.

\subsection{Spinning-down NS}
The difficulty of the fallback accretion model suggests that the
central engine of AT2017gfo should be long lasting, at least for
more than 10 days. The merger product can only be a supramassive or
even a permanently stable NS. Such an NS has long been suggested as
the central engine of GRBs (Dai \& Lu 1998a; 1998b; Zhang \&
M\'{e}sz\'{a}ros 2001; Dai et al. 2006), superluminous supernovae
(Kasen \& Bildsten 2010; Woolsy 2010) and also mergernovae (Yu et
al. 2013; Metzger \& Piro 2014; Yu et al. 2018). Different from Yu
et al. (2018) who invoked an NS to interpret the late-time emission
of AT2017gfo, this Letter invokes the NS power to interpret the
entire (blue and red) emission of AT2017gfo.

As analyzed by Yu et al. (2018), the surface dipolar magnetic filed
of the remnant NS of GW170817 cannot be very high if the NS spins at
a near-Keplerian frequency initially, because of the constraints
posed by the mergernova luminosity and timescale. A similar
constraint can be also derived from the multi-wavelength data (Ai et
al. 2018). In order to spin down the NS significantly, efficient
secular GW spindown is needed. In the GW-spindown-dominated regime,
the temporal evolution of the luminosity of the magnetic dipole
radiation of the NS, which is absorbed by the merger ejecta, can be
expressed as
\begin{eqnarray}
L_{\rm md}=L_{\rm md,i}\left(1+{t\over t_{\rm sd,gw}}\right)^{-1}
\end{eqnarray}
with
\begin{eqnarray}
L_{\rm md,i}=9.6\times10^{42} {\rm erg\,s^{-1}} \,R_{6}^6B_{12}^2 P_{\rm
i,-3}^{-4}\,
\end{eqnarray}
and
\begin{eqnarray}
t_{\rm sd,gw}=9.1\times 10^3 {\rm s}\,
\epsilon^{-2}_{-3}I_{45}^{-1}P_{\rm i,-3}^{4}\,
\end{eqnarray}
where $R$, $B$, $P_{\rm i}$, and $I$ are the radius, surface
magnetic field, initial spin period, and the moment of inertia of
the NS, respectively. The conventional notation $Q_{x}=Q/10^x$ is
adopted in cgs units.

\begin{figure}
\centering\resizebox{\hsize}{!}{\includegraphics{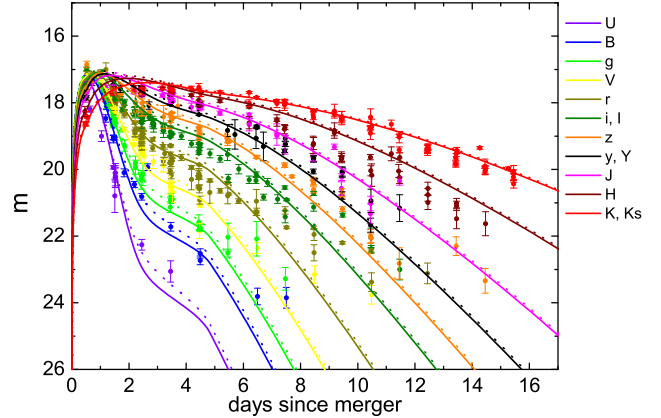}}
\caption{Same as in Fig.\ref{fitfb} but for energy injection from a
spinning-down NS. The fitting parameters are also shown in Table
\ref{table}. With small ejecta masses for both the polar and
equatorial components, heating due to radioactive decay is no longer
important for powering the observed emission during the entire
duration of the event.} \label{fitNS}
\end{figure}

A model fit to the multi-wavelength light curves of AT2017gfo with
energy injection of a low-$B$ NS is presented in Fig.\ref{fitNS},
which shows that both the early and late emission of AT2017gfo can
well be accounted for by this model. The model parameters are
collected in Table 1. In this model, one also needs two ejecta
components. The masses of the polar and equatorial ejecta are both
required to be on the order of $10^{-3}\rm M_{\odot}$, which
comfortably match the values of NS-NS merger simulations. The
heating due to radioactive decay is no longer important during the
entire emission episode, and the mergernova is dominantly powered by
energy injection from the NS. The GW-dominated spindown timescale is
adopted as $t_{\rm sd,g}=500\,\rm s$, which is inspired by the
extended emission or plateaus in SGRBs (Rowlinson et al. 2013;
L\"{u} et al. 2015). The initial magnetic dipole luminosity can be
then constrained to be $L_{\rm md,i}=3.4\times10^{44}\,\rm
erg\,s^{-1}$. For an initial Keplerian spin period $P_{\rm i}=1$ ms,
the ellipticity and the surface dipolar magnetic field of the NS can
be derived as
\begin{eqnarray}
\epsilon=0.0035
\end{eqnarray}
and
\begin{eqnarray}
B=3.4\times 10^{12}\rm G,
\end{eqnarray}
where a stellar radius of $R=1.2\times 10^{6}\rm\,cm$ and a moment
of inertia of $I=1.5\times10^{45}\rm g\,cm^{2}$ are adopted.
According to these parameters, the remnant NS during the mergernova
timescale should have a very high deformation but a relatively
normal poloidal surface magnetic field. These results are consistent
with the constraints recently given by Ai et al. (2018) and Yu et
al. (2018), and is also consistent with the requirement of
interpreting internal X-ray plateaus in short GRBs (Fan et al. 2013;
Gao et al. 2016). Specifically, the surface magnetic field found
here is about an order of magnitude higher than that found by Yu et
al. (2018), i.e., $\sim10^{12}$G versus $\sim10^{11}$G. This
relatively normal strength of the dipolar magnetic field could be
just an effective strength corresponding to the required spin-down
luminosity. The high ellipticity of the remnant NS strongly suggests
that its internal (probably toroidal) magnetic fields are ultrahigh,
i.e., the NS is a magnetar. Then, the surface magnetic field of the
NS could also be intrinsically much higher than $\sim 10^{12}G$,
which could however be significantly buried and/or exist in the form
of a multipolar filed (see Yu et al. 2018 for a detailed
disucssion). In any case, the existence of a long-lived NS is also
helpful to interpret the broadband afterglow of the event (Geng et
al. 2018).

\section{Summary and Conclusions}\label{conclusion}

Based on the traditional radioactivity-powered mergernova (kilonova)
model, we used the peak bolometric luminosity and peak emission time
of AT2017gfo to estimate the parameters of the merger ejecta and
obtained a large ejecta mass $\sim 0.04\,M_{\odot}$ and a low
opacity of $\kappa \sim 0.3\,\rm cm^{2}g^{-1}$. These are broadly
consistent with more detailed modeling by many authors. On the other
hand, we argue that this set of parameters is difficult to achieve
within the framework of NS-NS mergers without a long-lasting central
engine. In particular, the required $\kappa$ is too low even for
lanthanide-free ejecta. Even though a high $Y_{\rm e}$ is achieved,
 it is hard to reduce $\kappa$ to the desired value. Furthermore,
radioactive heating becomes inefficient with a high $Y_{\rm e}$, so
that an even larger ejecta mass is needed to achieve the desired
peak flux. These pose a problem for the standard model.

We then argue that a simple fix for this problem is to introduce a
central engine that can provide continuous energy injection to power
the mergernova emission. We find that a BH with fallback accretion
may power the bright early blue emission component of AT2017gfo, but
the late red component still needs radioactive heating of a massive
ejecta inconsistent with the numerical results of NS-NS mergers.
This only leaves us the option of having a long-lived, low-$B$ NS as
the central engine. We show that the multiband light curves of
AT2017gfo can well be reproduced by such an engine-driven mergernova
model, where both the opacity $\kappa$ and the ejected mass values
for both the blue and red components fall into the reasonable ranges
of known numerical simulations. This enhances the suggestion (Yu et
al. 2018) of a long-lived NS as the merger remnant of GW170817 (see
also Ai et al. 2018; Geng et al. 2018).

In addition to the engine-powered mergernova model discussed here,
we would also like to mention that Piro \& Kollmeier (2018)
suggested that the blue emission component of AT2017gfo can be
explained by the cooling of a shocked cocoon. The cocoon energy is
generated by the shock interaction between a relativistic GRB jet
and the surrounding envelope. In principle, this shock interaction
may be also regarded as a kind of energy injection. Furthermore,
Matsumoto et al. (2018) suggested that both the blue and red
emission components could be powered by energy injections from a jet
and X-rays, respectively (see Kisaka et al. 2016 for the original
suggestion of X-ray powered macronova, which is the kilonova and
mergernova discussed in this Letter). Murase et al. (2018) discussed
the possibility of differentiating the BH vs. NS engine using
high-energy emission from NS-NS merger systems.

\acknowledgements This work is supported by the National Basic
Research Program of China (973 Program, grant 2014CB845800), the
National Natural Science Foundation of China (grant No. 11473008 and
11573014), and the Self-Determined Research Funds of CCNU from the
colleges' basic research and operation of MOE. S.-Z. Li and L.-D.
Liu are supported by scholarships from the China Scholarship Council
(No. 201706770050 and No. 201706190127) to conduct research at the
University of Nevada, Las Vegas (UNLV).

\appendix
\section{The model}
Because the merger ejecta turns out to be mild relativistic, we
reduce the mergernova model (Yu et al. 2013) into the Newtonian
form. This semi-analytical model is also developed by involving the
density profile of the ejecta in order to give a better description
about the multiband light curves. Numerical simulations suggest that
the density profile of the dynamical ejecta cannot be fitted with
only one single power law (Piran et al. 2013). In addition, a
central engine may significantly modify the density profile into a
shell-like structure (Kasen et al. 2016).  Here we generally adopt a
broken power-law profile
\begin{subnumcases}
{\rho(r)=\rho_{0}}
\left({r\over R_{\rm ej}}\right)^{-\delta}, &$r\leqslant R_{\rm ej}$,\\
\left({r\over R_{\rm ej}}\right)^{-\zeta}, &$r> R_{\rm ej}$,
\end{subnumcases}
where $R_{\rm ej}$ is the radius of the main ejecta,
$\rho_{0}={(3-\delta)(\zeta-3)M_{\rm ej}/ 4\pi (\zeta-\delta)}{
R_{\rm ej}^3}$ is the density at $R_{\rm ej}$, and $M_{\rm ej}$ is
the mass. This is very similar to the density profile of a supernova
(Kasen et al. 2016). The difference here is that the
parameter $\delta$ is adopted as a negative value to represent a
shell-like structure. The density profile is expected to be shallow in the inner ejecta,
$r\leqslant R_{\rm ej}$, but become very steep in the outer ejecta,
$r> R_{\rm ej}$. So the parameter $\zeta$ is adopted as a relatively
large positive value. This profile avoids the infinite integral mass
problem, so that a sudden cut-off is not introduced.

The basic energy conservation equation is
\begin{eqnarray}
{dE\over dt}=-P{dV\over dt}-L_{\rm e}+L_{\rm in},
\label{dE/dt}
\end{eqnarray}
where $E$ is internal energy, $L_{\rm e}$ is emission luminosity,
$L_{\rm in}$ is energy injection rate, $V=4\pi R_{\rm ej}^{3}/3$ is
the main volume, and
\begin{equation}
P=\frac{E}{ 3V}- \frac{L_{\rm e}}{4\pi R_{\rm ej}^{2}c}
\label{Peff}
\end{equation}
is the effective (radiation dominated) pressure (with the second
term in Equation (\ref{Peff}) taking care of the leakage of
radiation pressure), and $P{dV}$ represents the energy lose due to
adiabatic expansion. Since the density out of $R_{\rm ej}$ drops
fast due to large $\zeta$, the velocity of ejecta is defined as
$v_{\rm ej}=dR_{\rm ej}/dt$. The initial velocity $v_{\rm ej,i}$ is
treated as a free parameter. Due to continuous energy injection, the
ejecta would be accelerated with
\begin{eqnarray}
{dv_{\rm ej}\over dt}={4\pi R_{\rm ej}^2 P\over M_{\rm ej}}.
\label{dv/dt}
\end{eqnarray}
Our treatment includes both central engine heating and radioactive
heating. The difference is that the former has the heating source at
the bottom, while the latter has heating throughout the ejecta. This
result is a small difference in calculating the optical depth of the
ejecta, with
\begin{equation}
\tau=\frac{(\zeta-\delta)\kappa\rho_{0}R_{\rm ej}}{(1-\delta)(\zeta-1)}
\end{equation}
denoting the integral optical depth from the bottom, which is relevant for central engine heating,
and
\begin{equation}
\tilde\tau \simeq \tau / \beta
\end{equation}
is relevant for radioactive heating, where $\beta \sim 3$ is a dimensionless parameter reflecting
the averaging effect in the ejecta (Arnett 1982; Metzger et al. 2010).

The emission luminosities for the central engine (subscript `c') and
radioactive (subscript `r') components can be written as (Kasen \&
Bildsten 2010; Yu et al. 2013)
\begin{subnumcases}
{L_{\rm e,c}=}
{E_{\rm c}c\over \tau R_{\rm ej}},& for $\tau\geqslant1$,\\
{E_{\rm c}c\over R_{\rm ej}},& for
$\tau<1$,
\end{subnumcases}
\begin{subnumcases}
{L_{\rm e,r}=}
{E_{\rm r}c\over \tilde{\tau} R_{\rm ej}},& for $\tilde\tau\geqslant1$,\\
{E_{\rm r}c\over R_{\rm ej}},& for
$\tilde\tau<1$,
\end{subnumcases}
where $c$ is the speed of light, and the total internal energy is
$E=E_{\rm c}+E_{\rm r}$.

For a central engine, the bolometric light curve would peak at the
photon diffusion timescale (Kasen \& Bildsten 2010; Yu et al. 2015),
i.e.,
\begin{eqnarray}
t_{\rm d}\approx(3\kappa M_{\rm ej}/4\pi v_{\rm ej}c)^{1/2},
\end{eqnarray}
if the density is uniform ($\delta=0$). When the density profile is
fully considered, this gives
\begin{eqnarray}
t_{\rm d}\approx\left[{{(3-\delta)(\zeta-3)\kappa M_{\rm
ej}}\over{(1-\delta)(\zeta-1)4\pi v_{\rm ej}c}}\right]^{1/2},
\end{eqnarray}
which depends on the parameters $\delta$ and $\zeta$.

The spectrum of emission is a blackbody when $\tau\geqslant1$ but
would be somewhat deviated when $\tau<1$. Near the peak, the ejecta
is still optically thick. One can define the effective temperature
$T_{\rm eff}=\sqrt[4]{L_{\rm e}/4\pi R_{\rm ph}^2\sigma}$, where
$R_{\rm ph}$ is the photosphere radius at which the optical depth
from $R_{\rm ph}$ to the outer edge of ejecta drops to
unity.\footnote{Notice that once a velocity is derived from a
spectrum analysis, the velocity should be the photosphere velocity
that is determined by $v_{\rm ph}=v_{\rm ej}{R_{\rm ph}/R_{\rm
ej}}$.} In our calculation, the spectrum is assumed to be a
blackbody all of the time. This is valid before and slightly after
the peak, but would not be valid at later times. When the ejecta
becomes optically thin, a modified photosphere radius $R_{\rm
ph}'=R_{\rm ej}-(V_{\rm ej}-V_{\rm ph})/4\pi R_{\rm ej}^{2}$ instead
of $R_{\rm ph}$ is used to calculate effective temperature. The
observed flux can be then given by
\begin{eqnarray}
F_{\nu}={2\pi h\nu^3\over c^2}{1\over {\rm exp}[h\nu/k_{\rm B}T_{\rm
eff}]-1}\left({R_{\rm ph}\over D}\right)^2,
\end{eqnarray}
where $\nu$ is frequency, $D$ is distance, and $h$ and $k_{\rm B}$
are the Plank constant and Boltzmann constant, respectively.

The injected energy from the central engine would not be fully
deposited into ejecta. For simplicity, an efficiency of
$\eta_{0}=0.5$ is adopted for both central engine heating and
radioactive heating. When the ejecta becomes optically thin, the
efficiency of central engine heating would be related to the optical
depth, i.e.,
\begin{eqnarray}
\eta_{\rm c}=\eta_{0}(1-e^{-A\tau}),
\end{eqnarray}
where the parameter $A$ represents a characteristic efficiency decay rate. Similarly,
the efficiency corresponding to radioactive heating is given by
\begin{eqnarray}
\eta_{\rm r}=\eta_{0}(1-e^{-A\tilde{\tau}}).
\end{eqnarray}
The total energy injection from both central engine and radioactive
heating can be expressed as
\begin{eqnarray}
L_{\rm in}=\eta_{\rm c} L_{\rm c}+\eta_{\rm r}\varepsilon M_{\rm
ej},
\end{eqnarray}
where $L_{\rm c}$ is the central engine and $\varepsilon$ is the
heating rate from radioactive decay. So, once $L_{\rm c}$ and
$\varepsilon$ are given, the differential equations can be solved.

When two heating sources are invoked, separate differential
equations analogous to Equations (\ref{dE/dt}-\ref{dv/dt}) for both
the central engine heating and radioactive heating are solved.

In the above treatment, an isotropic ejecta (with solid angle
$4\pi$) is assumed. When fitting the data of AT2017gfo, we have
divided the ejecta into two components, a blue polar component with
solid angle $\Omega_{\rm p}$ and a red equatorial component with
solid angle $\Omega_{\rm e}$. In our model fitting, both components
have a solid angle $2\pi$, and the relevant ejecta parameters
($\kappa$ and $M$) are fitted separately (see Table 1).


\begin{thebibliography}{99}

\bibitem[Abbott et al.(2016)]{2016PhRvL.116f1102A} Abbott, B.~P., Abbott, R., Abbott, T.~D., et al.\ 2016, Physical Review Letters, 116, 061102

\bibitem[Abbott et al.(2017a)]{2017PhRvL.119p1101A} Abbott, B.~P., Abbott, R., Abbott, T.~D., et al.\ 2017a, Physical Review Letters, 119, 161101
\bibitem[Abbott et al.(2017b)]{2017ApJ...848L..13A} Abbott, B.~P., Abbott, R., Abbott, T.~D., et al.\ 2017b, \apjl, 848, L13

\bibitem[Ai et al.(2018)]{2018ApJ...860...57A} Ai, S., Gao, H., Dai, Z.-G., et al.\ 2018, \apj, 860, 57

\bibitem[Arcavi et al.(2017)]{2017Natur.551...64A} Arcavi, I., Hosseinzadeh, G., Howell, D.~A., et al.\ 2017, \nat, 551, 64

\bibitem[Arnett(1982)]{1982ApJ...253..785A} Arnett, W.~D.\ 1982, \apj, 253, 785

\bibitem[Banik \& Bandyopadhyay(2017)]{2017arXiv171209760B} Banik, S., \& Bandyopadhyay, D.\ 2017, arXiv:1712.09760

\bibitem[Barnes \& Kasen(2013)]{2013ApJ...775...18B} Barnes, J., \& Kasen, D.\ 2013, \apj, 775, 18

\bibitem[Barnes et al.(2016)]{2016ApJ...829..110B} Barnes, J., Kasen, D., Wu, M.-R., \& Mart{\'{\i}}nez-Pinedo, G.\ 2016, \apj, 829, 110

\bibitem[Bauswein et al.(2013)]{2013ApJ...773...78B} Bauswein, A., Goriely, S., \& Janka, H.-T.\ 2013, \apj, 773, 78

\bibitem[Coulter et al.(2017)]{2017Sci...358.1556C} Coulter, D.~A., Foley, R.~J., Kilpatrick, C.~D., et al.\ 2017, Science, 358, 1556

\bibitem[Cowperthwaite et al.(2017)]{2017ApJ...848L..17C} Cowperthwaite, P.~S., Berger, E., Villar, V.~A., et al.\ 2017, \apjl, 848, L17

\bibitem[Dai \& Lu(1998a)]{1998A&A...333L..87D} Dai, Z.~G., \& Lu, T.\ 1998a, \aap, 333, L87

\bibitem[Dai \& Lu(1998b)]{1998PhRvL..81.4301D} Dai, Z.~G., \& Lu, T.\ 1998b, Physical Review Letters, 81, 4301

\bibitem[Dai et al.(2006)]{} Dai, Z.~G., Wang, X.-Y., Wu, X.-F., \& Zhang, B.\ 2006, Science, 311, 1127

\bibitem[Dessart et al.(2009)]{2009ApJ...690.1681D} Dessart, L., Ott, C.~D., Burrows, A., Rosswog, S., \& Livne, E.\ 2009, \apj, 690, 1681

\bibitem[Fan et al.(2013)]{} Fan, Y.-Z., Wu, X.-F., \& Wei, D.-M. 2013, \prd, 88, 067304

\bibitem[Fern{\'a}ndez et al.(2015)]{2015MNRAS.446..750F} Fern{\'a}ndez, R., Kasen, D., Metzger, B.~D., \& Quataert, E.\ 2015, \mnras, 446, 750

\bibitem[Fern{\'a}ndez \& Metzger(2016)]{2016ARNPS..66...23F} Fern{\'a}ndez, R., \& Metzger, B.~D.\ 2016, Annual Review of Nuclear and Particle Science, 66, 23

\bibitem[Fern{\'a}ndez \& Metzger(2013)]{2013MNRAS.435..502F} Fern{\'a}ndez, R., \& Metzger, B.~D.\ 2013, \mnras, 435, 502

\bibitem[Foucart et al.(2013)]{2013PhRvD..87h4006F} Foucart, F., Deaton, M.~B., Duez, M.~D., et al.\ 2013, \prd, 87, 084006


\bibitem[Gao et al.(2016)]{2016PhRvD..93d4065G} Gao, H., Zhang, B., \& L{\"u}, H.-J.\ 2016, \prd, 93, 044065

\bibitem[Geng et al.(2018)]{2018arXiv180307219G} Geng, J.-J., Dai, Z.-G., Huang, Y.-F., et al.\ 2018, \apj, 856, L33

\bibitem[Goldstein et al.(2017)]{2017ApJ...848L..14G} Goldstein, A., Veres, P., Burns, E., et al.\ 2017, \apjl, 848, L14

\bibitem[Grossman et al.(2014)]{2014MNRAS.439..757G} Grossman, D., Korobkin, O., Rosswog, S., \& Piran, T.\ 2014, \mnras, 439, 757

\bibitem[Hotokezaka et al.(2013)]{2013PhRvD..87b4001H} Hotokezaka, K., Kiuchi, K., Kyutoku, K., et al.\ 2013, \prd, 87, 024001

\bibitem[Just et al.(2015)]{2015MNRAS.448..541J} Just, O., Bauswein, A., Ardevol Pulpillo, R., Goriely, S., \& Janka, H.-T.\ 2015, \mnras, 448, 541

\bibitem[Karp et al.(1977)]{1977ApJ...214..161K} Karp, A.~H., Lasher, G., Chan, K.~L., \& Salpeter, E.~E.\ 1977, \apj, 214, 161

\bibitem[Kasen et al.(2013)]{2013ApJ...774...25K} Kasen, D., Badnell, N.~R., \& Barnes, J.\ 2013, \apj, 774, 25

\bibitem[Kasen \& Bildsten(2010)]{2010ApJ...717..245K} Kasen, D., \& Bildsten, L.\ 2010, \apj, 717, 245

\bibitem[Kasen et al.(2015)]{2015MNRAS.450.1777K} Kasen, D., Fern{\'a}ndez, R., \& Metzger, B.~D.\ 2015, \mnras, 450, 1777

\bibitem[Kasen et al.(2016)]{2016ApJ...821...36K} Kasen, D., Metzger, B.~D., \& Bildsten, L.\ 2016, \apj, 821, 36

\bibitem[Kasen et al.(2017)]{2017Natur.551...80K} Kasen, D., Metzger, B., Barnes, J., Quataert, E., \& Ramirez-Ruiz, E.\ 2017, \nat, 551, 80

\bibitem[Kisaka et al.(2016)]{2016ApJ...818..104K} Kisaka,
S., Ioka, K., \& Nakar, E.\ 2016, \apj, 818, 104

\bibitem[Korobkin et al.(2012)]{2012MNRAS.426.1940K} Korobkin, O., Rosswog, S., Arcones, A., \& Winteler, C.\ 2012, \mnras, 426, 1940

\bibitem[L{\"u} et al.(2015)]{2015ApJ...805...89L} L{\"u}, H.-J., Zhang, B., Lei, W.-H., Li, Y., \& Lasky, P.~D.\ 2015, \apj, 805, 89

\bibitem[Li \& Paczy{\'n}ski(1998)]{1998ApJ...507L..59L} Li, L.-X., \& Paczy{\'n}ski, B.\ 1998, \apjl, 507, L59

\bibitem[Li \& Yu(2016)]{2016ApJ...819..120L} Li, S.-Z., \& Yu, Y.-W.\ 2016, \apj, 819, 120

\bibitem[Lippuner \& Roberts(2015)]{2015ApJ...815...82L} Lippuner, J., \& Roberts, L.~F.\ 2015, \apj, 815, 82

\bibitem[Lipunov et al.(2017)]{2017MNRAS.465.3656L} Lipunov, V.~M., Kornilov, V., Gorbovskoy, E., et al.\ 2017, \mnras, 465, 3656

\bibitem[Ma et al.(2018)]{2018ApJ...852L...5M} Ma, S.-B., Lei, W.-H., Gao, H., et al.\ 2018, \apjl, 852, L5

\bibitem[Matsumoto et al.(2018)]{2018arXiv180207732M} Matsumoto, T., Ioka, K., Kisaka, S., \& Nakar, E.\ 2018, arXiv:1802.07732

\bibitem[Metzger et al.(2010)]{2010MNRAS.406.2650M} Metzger, B.~D., Mart{\'{\i}}nez-Pinedo, G., Darbha, S., et al.\ 2010, \mnras, 406, 2650

\bibitem[Metzger(2017a)]{2017LRR....20....3M} Metzger, B.~D.\ 2017a, Living Reviews in Relativity, 20, 3


\bibitem[Metzger \& Fern{\'a}ndez(2014)]{2014MNRAS.441.3444M} Metzger, B.~D., \& Fern{\'a}ndez, R.\ 2014, \mnras, 441, 3444

\bibitem[Metzger \& Piro(2014)]{2014MNRAS.439.3916M} Metzger, B.~D., \& Piro, A.~L.\ 2014, \mnras, 439, 3916


\bibitem[Metzger et al.(2018)]{2018ApJ...856..101M} Metzger, B.~D., Thompson, T.~A., \& Quataert, E.\ 2018, \apj, 856, 101

\bibitem[Murase et al.(2018)]{2018ApJ...854...60M}  Murase,
K., Toomey, M.~W., Fang, K., et al.\ 2018, \apj, 854, 60

\bibitem[Nicholl et al.(2017)]{2017ApJ...848L..18N} Nicholl, M., Berger, E., Kasen, D., et al.\ 2017, \apjl, 848, L18

\bibitem[Perego et al.(2014)]{2014MNRAS.443.3134P} Perego, A., Rosswog, S., Cabez{\'o}n, R.~M., et al.\ 2014, \mnras, 443, 3134

\bibitem[Pinto \& Eastman(2000)]{2000ApJ...530..757P} Pinto, P.~A., \& Eastman, R.~G.\ 2000, \apj, 530, 757
\bibitem[Piran et al.(2013)]{2013MNRAS.430.2121P} Piran, T., Nakar, E., \& Rosswog, S.\ 2013, \mnras, 430, 2121

\bibitem[Piro \& Kollmeier(2018)]{2018ApJ...855..103P} Piro, A.~L., \& Kollmeier, J.~A.\ 2018, \apj, 855, 103

\bibitem[Radice et al.(2018)]{2018arXiv180310865R} Radice, D., Perego, A., Bernuzzi, S., \& Zhang, B.\ 2018, arXiv:1803.10865

\bibitem[Richers et al.(2015)]{2015ApJ...813...38R} Richers, S., Kasen, D., O'Connor, E., Fern{\'a}ndez, R., \& Ott, C.~D.\ 2015, \apj, 813, 38

\bibitem[Rosswog(2013)]{2013RSPTA.37120272R} Rosswog, S.\ 2013, Philosophical Transactions of the Royal Society of London Series A, 371, 20120272

\bibitem[Rosswog et al.(2017)]{2017CQGra..34j4001R} Rosswog, S., Feindt, U., Korobkin, O., et al.\ 2017, Classical and Quantum Gravity, 34, 104001

\bibitem[Rosswog(2007)]{2007MNRAS.376L..48R} Rosswog, S.\ 2007, \mnras, 376, L48

\bibitem[Rowlinson et al.(2013)]{2013MNRAS.430.1061R} Rowlinson, A., O'Brien, P.~T., Metzger, B.~D., Tanvir, N.~R., \& Levan, A.~J.\ 2013, \mnras, 430, 1061

\bibitem[Savchenko et al.(2017)]{2017ApJ...848L..15S} Savchenko, V., Ferrigno, C., Kuulkers, E., et al.\ 2017, \apjl, 848, L15

\bibitem[Shibata et al.(2017)]{2017PhRvD..95h3005S} Shibata, M., Kiuchi, K., \& Sekiguchi, Y.-i.\ 2017, \prd, 95, 083005

\bibitem[Siegel \& Metzger(2018)]{2018ApJ...858...52S} Siegel, D.~M., \& Metzger, B.~D.\ 2018, \apj, 858, 52

\bibitem[Smartt et al.(2017)]{2017Natur.551...75S} Smartt, S.~J., Chen, T.-W., Jerkstrand, A., et al.\ 2017, \nat, 551, 75

\bibitem[Song et al.(2018)]{2018MNRAS.477.2173S} Song, C.-Y., Liu, T., \& Li, A.\ 2018, \mnras, 477, 2173

\bibitem[Tanaka \& Hotokezaka(2013)]{2013ApJ...775..113T} Tanaka, M., \& Hotokezaka, K.\ 2013, \apj, 775, 113

\bibitem[Tanaka et al.(2018)]{2018ApJ...852..109T} Tanaka, M., Kato, D., Gaigalas, G., et al.\ 2018, \apj, 852, 109

\bibitem[Tanaka et al.(2017)]{2017PASJ...69..102T} Tanaka, M., Utsumi, Y., Mazzali, P.~A., et al.\ 2017, \pasj, 69, 102

\bibitem[Tanvir et al.(2017)]{2017ApJ...848L..27T} Tanvir, N.~R., Levan, A.~J., Gonz{\'a}lez-Fern{\'a}ndez, C., et al.\ 2017, \apjl, 848, L27

\bibitem[Troja et al.(2017)]{2017Natur.551...71T} Troja, E., Piro, L., van Eerten, H., et al.\ 2017, \nat, 551, 71

\bibitem[Valenti et al.(2017)]{2017ApJ...848L..24V} Valenti, S., David, Sand, J., et al.\ 2017, \apjl, 848, L24

\bibitem[Villar et al.(2017)]{2017ApJ...851L..21V} Villar, V.~A., Guillochon, J., Berger, E., et al.\ 2017, \apjl, 851, L21

\bibitem[Wanajo et al.(2014)]{2014ApJ...789L..39W} Wanajo, S., Sekiguchi, Y., Nishimura, N., et al.\ 2014, \apjl, 789, L39

\bibitem[Waxman et al.(2017)]{2017arXiv171109638W} Waxman, E., Ofek, E., Kushnir, D., \& Gal-Yam, A.\ 2017, arXiv:1711.09638

\bibitem[Woosley(2010)]{2010ApJ...719L.204W} Woosley, S.~E.\ 2010, \apjl, 719, L204

\bibitem[Yu et al.(2013)]{2013ApJ...776L..40Y} Yu, Y.-W., Zhang, B., \& Gao, H.\ 2013, \apjl, 776, L40

\bibitem[Yu et al.(2015)]{2015ApJ...806L...6Y} Yu, Y.-W., Li, S.-Z., \& Dai, Z.-G.\ 2015, \apjl, 806, L6

\bibitem[Yu et al.(2018)]{2017arXiv171101898Y} Yu, Y.-W., Liu, L.-D., \& Dai, Z.-G.\ 2018, \apj, in press, arXiv:1711.01898

\bibitem[Zhang \& M{\'e}sz{\'a}ros(2001)]{2001ApJ...552L..35Z} Zhang, B., \& M{\'e}sz{\'a}ros, P.\ 2001, \apjl, 552, L35

\bibitem[Zhang et al.(2018)]{zhang18} Zhang, B.-B., Zhang, B., Sun, H. et al. \ 2018, Nature Communications, 9, 447

\end{thebibliography}
\end{document}